\documentclass[twocolumn,preprintnumbers,amsmath,amssymb,secnumarabic,amssymb, aps, prd,showpacs]{revtex4-1}
\usepackage{graphicx}
\usepackage{dcolumn}
\usepackage{bm}
\usepackage{appendix}
\newcommand{\ds}{\displaystyle}

\newcommand{\be}{\begin{equation}}
\newcommand{\ee}{\end{equation}}
\newcommand{\beq}{\begin{eqnarray}}
\newcommand{\eeq}{\end{eqnarray}}

\newcommand{\w}{\omega}
\newcommand{\W}{\Omega}
\newcommand{\g}{\gamma}
\newcommand{\G}{\Gamma}

\newcommand{\s}{\sigma}

\newcommand{\bnn}{\begin{eqnarray*}}
\newcommand{\enn}{\end{eqnarray*}}
\begin{document}
\title{Coherent Excitation of a Two-Level Atom driven by a far off-resonant Classical Field: Analytical Solutions}
\author{Pankaj K. Jha$^{1,*}$}
\author{Yuri V. Rostovtsev$^{1,2}$}%
\affiliation{$^{1}$Institute for Quantum Science and Engineering and Department of Physics, Texas A\&M University, TX 77843.\\
$^{2}$Department of Physics, University of North Texas, Denton TX 76203}
\begin{abstract}
 We present an analytical treatment of coherent excitation of a Two-Level Atom driven by a far-off resonant classical field. A class of pulse envelope is obtained for which this problem is exactly solvable. The solutions are given in terms of Heun function which is a generalization of the Hypergeometric function. The degeneracy of Heun to Hypergeometric equation can give all the exactly solvable pulse shapes of Gauss Hypergeometric form, from the generalized pulse shape obtained here. We discuss the application of the results obtained to the generation of XUV.
\end{abstract}
\date{\today}
\maketitle
\section{Introduction}
 The two-level system (TLS)[1-3] is a very rich and useful model that helps to understand physics of many problems ranging from interaction with electromagnetic fields to level-crossing[4-6]. For example, interaction of a beam of atoms in Stern-Gerlach apparatus[7], and Bloch-Siegert shift[8] can be understood using TLS. Recently TLS has been extensively studied as a quantum bit (qubit) for quantum information theory[9]. Two-level atom (TLA) description is valid if the two atomic levels involved are resonant or nearly resonant with the driving field, while all other levels are highly detuned. TLS  can be realized exactly for a spin-1/2 system, and, approximately, for a multi-level system in a magnetic field when all other magnetic sub-levels are detuned far-off resonance.

When the frequency of the driving field is in resonance with the atomic transition frequency, the Schr$\ddot{o}$dinger equation for the time evolution of state amplitudes is exactly solvable for any time dependence of the field $\Omega(t)$. The transition probability is given as
\begin{equation}
 p=\mbox{sin}^{2}(A/2), \quad A=\int_{-\infty}^{\infty}{\Omega(\tilde{t})}d\tilde{t}.
 \end{equation}
Here $A$ is the area of the pulse envelope. Interestingly this transition probability vanishes when $A$ is an even integer multiple of $\pi$ (CPR). For odd integer multiple of $\pi$ we get complete population inversion (CPI) while half-integer multiple of $\pi$ gives equal coherent superposition of the initial and the final states. Several exactly solvable models for the TLS have been proposed in the past[10-24] where solutions to the Schr$\ddot{o}$dinger equation are expressed in terms of known functions like Hypergeomteric functions. Several approximate solutions have also been proposed based on perturbation theory and the adiabatic approximation[25,26].

Recently, the topic has been in a focus of research related to generation of short wavelength radiation[27,28]. A two-level atomic system under the action of a far-off resonance strong pulse of laser radiation has been considered and it has been shown that such pulses can excite remarkable coherence on high frequency far-detuned transitions; and this coherence can be used for efficient generation of UV and soft X-ray (XUV) radiation[28].

To describe excited coherence, we are interested to understand the mechanism of breaking adiabaticity that leads to excited coherence in the system when the laser pulse has already passed. Thus we are interested going beyond  classical electrodynamics[29]. Indeed, an electric field causes polarization of dielectrics is given by
\begin{equation}
P(t,r)=\int_{-\infty}^{t}dt'\chi(t-t'){\cal E}(t',r),
\end{equation}
where $\chi(\tau)$ is the dielectric response function. It is important to note that once the field is removed, the polarization adiabatically returns to practically zero. Breaking of adiabaticity is especially difficult when the frequency of the applied field is far from the atomic resonance. Finding exact analytical solutions for such a problem will not only supplement numerical simulations but will also be useful in understanding the underlying physics.

In this paper, using a proper variable transformation,  we find a class of pulse $\Omega(t)$  for which the Schr$\ddot{o}$dinger equation for the time evolution of the state amplitudes can be transformed into the well known Heun equation.  The solutions are given in terms of the Heun function which is a generalization of the Hypergeometric function. Using the degeneracy of Heun to Hypergeometric equation, Bambini-Berman model[21] can be generalized to this model.

The paper is organized as follows. In section 2, we briefly describe our system and obtain the equation of motion for the state amplitudes. In section 3, we present the exact solution of the problem in terms of the local Heun solutions H$l$. It is well established that the Heun equation reduces to the Gauss Hypergeometric equation in several ways so we discuss this degeneracy briefly.  We also discuss one of the confluent cases of the Heun Equation i.e the Confluent Heun Equation and find the exact solutions. In section 4, we give some specific examples of the pulses for which we have found solutions. Pulse shapes are asymmetric in time except the Rosen-Zener pulse. The Hyperbolic secant (Rosen-Zener Model), generalized Rosen-Zener (Bambini-Berman) Model are included in this class $\Omega(t)$ as a special case. We also give a new model for a Smooth Box Pulse which takes care of non-analyticity at the edges by introducing a parameter $\delta$. By modulating this parameter we can modulate the box width. In section 5, we discuss the application of the results obtained here to the generation of XUV. We also estimate the level of the XUV field that can be generated by using the excited coherence.
\section{ Two-Level Atom: Equation of Motion}
The equation of motion for the probability amplitudes for the states $|a\rangle$ and $|b\rangle$ (see Fig1(a)) of a Two Level Atom (TLA) interacting with a classical field is given as
\begin{subequations}\label{TLA1}
\begin{align}
\dot{C}_{a}&= i\frac{\wp{\cal E}(t)}{\hbar}\mbox{cos}(\nu t) e^{i\omega t}C_{b},\label{second}\\
\dot{C}_{b}&= i\frac{\wp^{*}{\cal E}(t)}{\hbar}\mbox{cos}(\nu t)e^{-i\omega t}C_{a},
\end{align}
\end{subequations}
\begin{figure}[b]
  \includegraphics[height=3.5cm,width=3.2cm,angle=0]{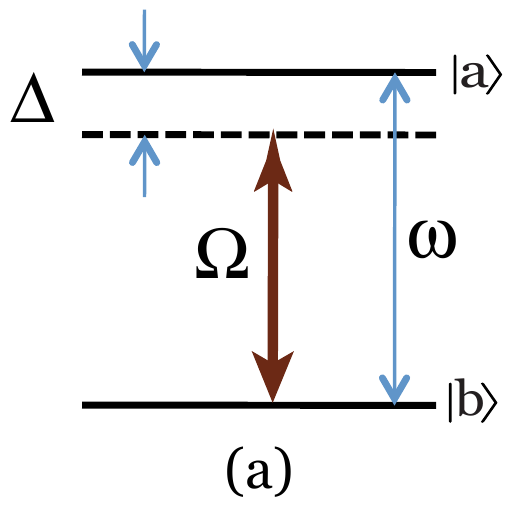}
  \includegraphics[height=3.5cm,width=4.7cm,angle=0]{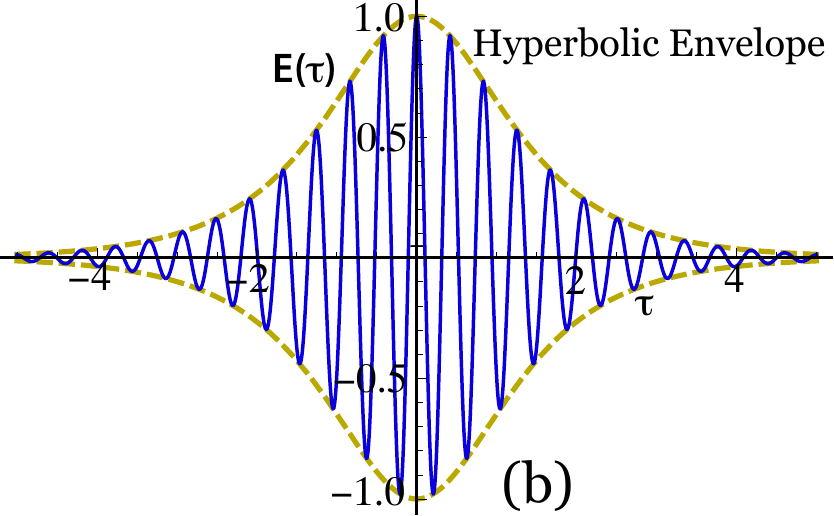}
  \caption{(Color online) (a)Two-level atomic system, atomic transition frequency $\omega=\omega_{a}-\omega_{b}$, detuning $ \Delta = \omega -\nu$ and Rabi frequency $\Omega(t)=\wp{\cal E}(t)/2\hbar$. (b)  Classical electromagnetic field E(t)$={\mbox{sech}(\alpha}t)\mbox{cos}(\nu t)$}
\end{figure}
where $\hbar\omega$ is the energy difference between two levels, $\wp$ is the atomic dipole moment; $E(t)={\cal E}(t)\mbox{cos}\nu t$ (see Fig1(b)). In the Rotating Wave Approximation (RWA) we replace $\mbox{cos}(\nu t)e^{\pm i\omega t}\rightarrow e^{\pm i\Delta}/2$ where $\Delta=\omega-\nu$, is detuning from resonance. Introducing $\Omega(t)=\wp{\cal E}(t)/2\hbar$[30], Eq.(\ref{TLA1}) reduces to
\begin{subequations}\label{TLA2}
\begin{align}
\dot{C}_{a}&= i\Omega(t) e^{i\Delta t}C_{b},\label{second}\\
\dot{C}_{b}&= i\Omega^{*}(t)e^{-i\Delta t}C_{a},
\end{align}
\end{subequations}
which have an integral of motion $|C_{a}|^{2} + |C_{b}|^{2} = 1$[31]. There are a variety  of ways to approach the problem of solving for $C_{a}(t)$ . One method is to define $f(t)=C_{a}(t)/C_{b}(t)$. For the function $f(t)$, Eq.(\ref{TLA2}) yields the following Riccati Equation[28]
\begin{equation}
\dot{f}+i\Omega^{*}(t)e^{-i\Delta t}f^{2}-i\Omega(t)e^{i\Delta t}=0.
\end{equation}
Then $|C_{a}(t|=|f(t)| / \sqrt{1+|f(t)|^{2}}$. Alternatively, we can get a second order linear differential equation for $C_{a}(t)$, from Eq.(\ref{TLA2})
\begin{equation}\label{L4}
\ddot{C}_{a}(t)-\left[i\Delta + \frac{\dot{\Omega}}{\Omega}\right]\dot{C}_{a}(t)+|\Omega|^{2}C_{a}(t)=0.
\end{equation}
The general solution for Eq(\ref{L4}) has not been found yet, however there are solutions for several cases in terms of special functions.
\noindent To find a solution for Eq.(\ref{L4}) we introduce a new variable
\begin{equation}
\varphi=\varphi(\tau),
\end{equation}
subject to the condition that $\varphi (\tau)$ is real, positive and monotonic function of $\tau$ and $\varphi_{0}\leq \varphi \leq \varphi_{1}$.
In terms of the variable $\varphi$ and the dimensionless parameters
\begin{equation}\label{L5}
\tau = \alpha t, \quad \beta = \frac{\Delta}{\alpha}, \quad \gamma = \frac{\Omega_{0}}{\alpha},
\end{equation}
one may write Eq.(\ref{L4}), for real $\xi(\tau)$ in the form
\begin{equation}\label{L7}
C_{a}^{''}+\left[\frac{\ddot{\varphi}/\dot{\varphi}-i\beta-\dot{\xi}/\xi}{\dot{\varphi}}\right]C_{a}^{'}+\frac{\gamma^{2}\xi^{2}}{\dot{\varphi}^{2}}C_{a}=0,
\end{equation}
where a prime indicates differentiation with respect to $\varphi$ and $\Omega(\tau)=\gamma \xi(\tau)$. Let us determine the condition under which Eq.(\ref{L7}) has the form
\begin{equation}\label{L8}
C_{a}^{''}(\varphi)+P(\varphi)C_{a}^{'}(\varphi)+Q(\varphi)C_{a}(\varphi)=0.
\end{equation}
Using Eq.(\ref{L7},\ref{L8}) and some trivial alebra we get,
\begin{equation}
\tau=-\frac{1}{i\beta}\int\left(P+\frac{Q^{'}}{2Q} \right)d\varphi.
\end{equation}
\section{ Heun Equation}
Bambini-Berman studied the case in which Eq.(\ref{L8}) has the form of a Gauss Hypergeometric equation which includes Rosen-Zener Model as a special case. Now let us consider when Eq.(\ref{L8}) is of the form of Heun equation [32,33] with the independent variable $\varphi$.
\begin{equation}\label{L10}
\frac{d^{2}C_{a}}{d\varphi^{2}}+\left(\frac{u}{\varphi}+\frac{v}{\varphi-1}+\frac{w}{\varphi-c}\right)\frac{dC_{a}}{d\varphi}+
\frac{(ab\varphi-q)C_{a}}{\varphi(\varphi-1)(\varphi-c)}=0,
\end{equation}
where \textit{a,b,c,q,u,v,w} are parameters with $c\ne 0,1$.$(c>1)$. The parameters are constrained, by the general theory of Fuchsian equations, as
\begin{equation}
u+v+w=a+b+1.
\end{equation}
 From Eq.(\ref{L10}) and Eq.(\ref{L8}) and some algebra we get
\begin{equation}\label{L13}
\dot{\varphi}=2\varphi(1-\varphi)/(\mu+\lambda\varphi).
\end{equation}
Equivalently the parameters of the Heun Equation Eq.(\ref{L10}) are given as
\begin{equation}
\begin{split}
u= \frac{1}{2}-\frac{i\beta\mu}{2},& \quad v= \frac{1}{2}+\frac{i\beta(\lambda +\mu)}{2},\quad w= \frac{1}{2},\\
& \quad a= 0, \quad b=\frac{1}{2}-\frac{i\beta \lambda}{2}.
\end{split}
\end{equation}
For $\varphi(\tau)$ to be a monotonically increasing function of $\tau$, $\dot{\varphi}$ must be real and positive i.e
$\mu>0, \lambda/\mu>-1.$
\noindent The time variable $\tau$ as a function of $\varphi$ is obtained by integrating Eq.(\ref{L13}) which gives,
\begin{equation}\label{L14}
2\tau=\mbox{ln}[\varphi^{\mu}/(1-\varphi)^{\mu+\lambda}].
\end{equation}
\begin{figure*}
 \includegraphics[height=6cm,width=0.45\textwidth]{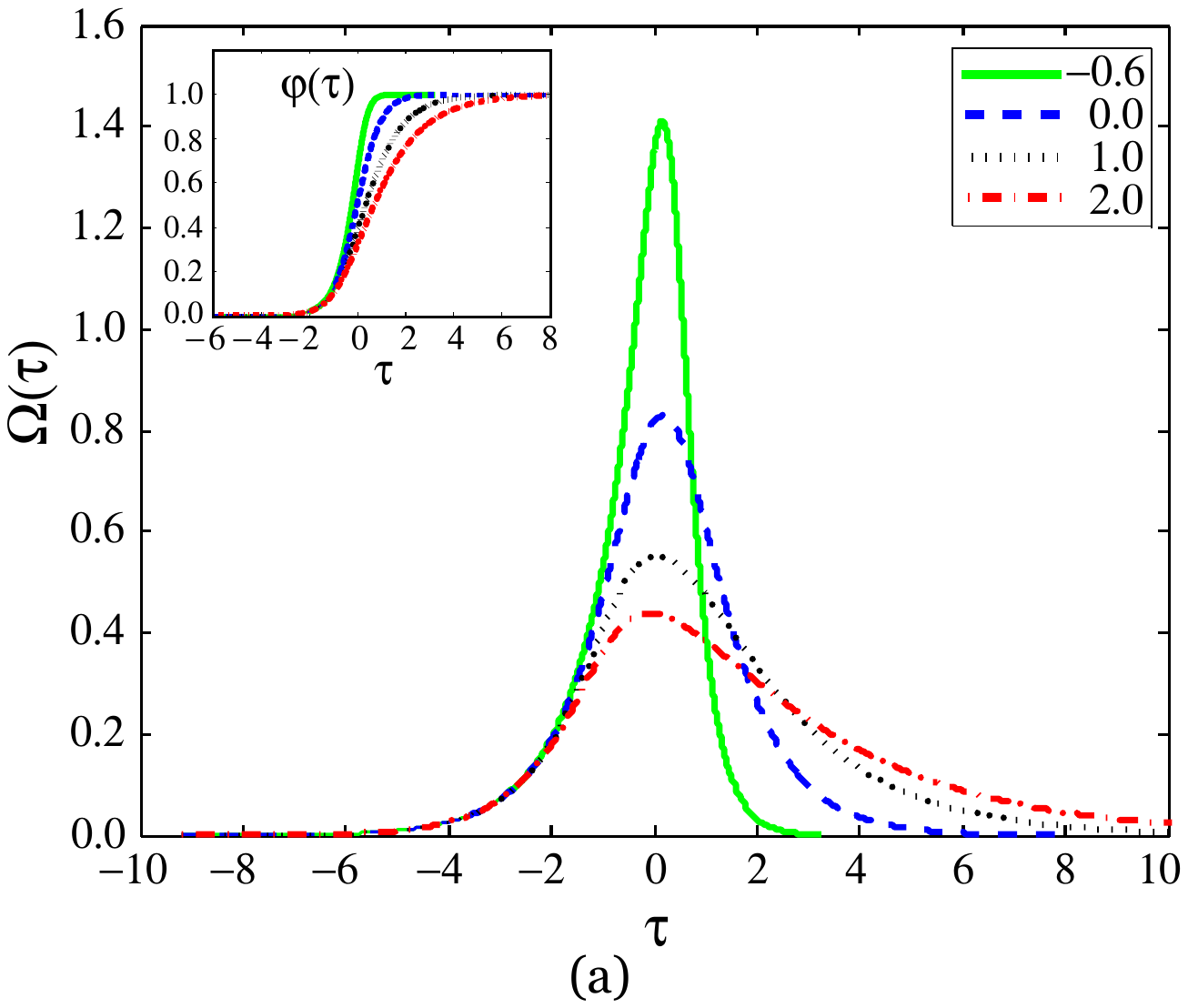}
\includegraphics[height=6cm,width=0.45\textwidth,angle=0]{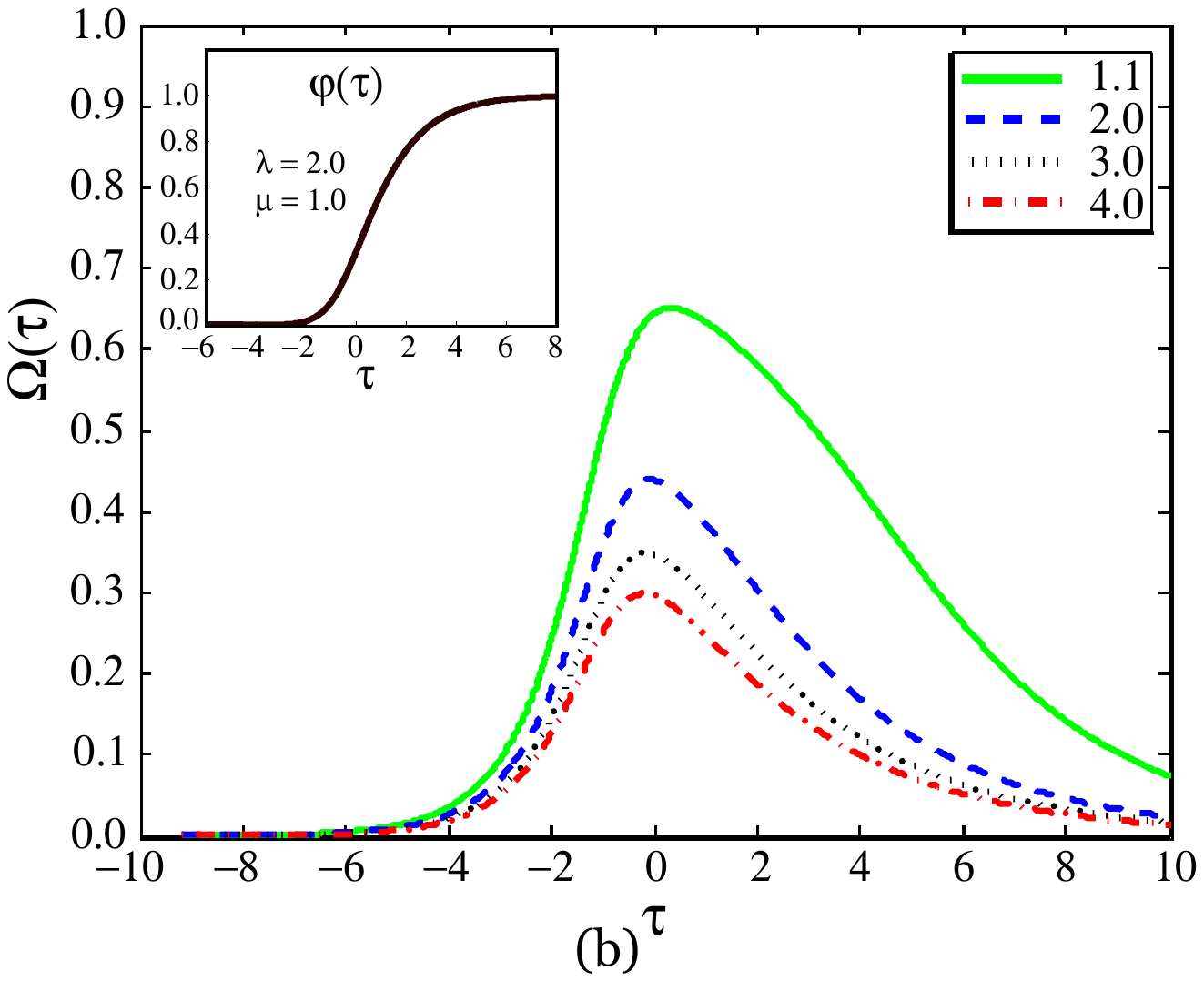}
 \caption{(Color online) Pulse shapes given by Eq(18). (a)  Pulse shapes with varying $\lambda$ and $ c=2, q=-1,ab=0$. (b) Pulse shapes with varying $c$ and $\lambda=2, q=-1,ab=0.$}
\end{figure*}
The general solution for Eq.(\ref{L10}), which has regular singularity at $\varphi=0$ is given in terms of the Heun local solutions, H\textit{l}$(\varphi)$ as,
\begin{align}\label{L15}
  \begin{split}
   & C_{a}=\mbox{P}_{1}\varphi^{1-u}\mbox{H}l[c,q+(1-u)((c-1)v+a+b-u+1);\\
    & a-u+1,b-u+1,2-u,v;\varphi]+\mbox{P}_{2}\mbox{H}l\left[c,q;a,b,u,v;\varphi\right],
  \end{split}
\end{align}
where the constants, $\mbox{P}_{1} , \mbox{P}_{2}$ can be found using the initial conditions of the system. In the limit $\tau \rightarrow \infty $, the population left in the level $|a\rangle$ can be obtained by substituting $\varphi \rightarrow 1$ in Eq.(\ref{L15}). The form of the pulse can be obtained by equating Eq.(\ref{L7}) and Eq.(\ref{L10}) which gives
\begin{equation}\label{Pulse}
\Omega(\tau)=\left[\frac{4\varphi(1-\varphi)(ab\varphi-q)}{(c-\varphi)}\right]^{1/2}\left(\frac{1}{\mu+\lambda\varphi}\right),
\end{equation}
where $\varphi(\tau)$ is given by Eq.(\ref{L14})[34]. In Fig (2)  we have plotted the pulse envelopes'  of the classical field, given by Eq.(\ref{Pulse}), for which the two-level atom problem can be exactly solved. They also show the effect of the asymmetric parameters $\lambda$ and $ab$ respectively, for $\mu=1$, on the symmetry of the shapes. Pulse shapes showing the effects of other parameters can also be plotted easily from Eq(\ref{Pulse}).

There are three kinds of solutions to the Heun equation Eq.(\ref{L10}). Local Solutions H$\it{l}$, Heun functions H$f$ and Heun Polynomials H$p$[35-37].
The series solution Eq.(\ref{L15}) is written as [33]
\begin{equation}\label{L16}
\mbox{H}l\left[c,q;a,b,u,v;\varphi\right]=\sum_{j=0}^{\infty}s_{j}\varphi^{j}=1+\frac{q}{uc}\varphi+\sum_{j=2}^{\infty}s_{j}\varphi^{j},
\end{equation}
where $s_{j}$ obeys the three term recursion relation
\begin{align}\label{L17}
  \begin{split}
    &(j-1+a)(j-1+b)s_{j-1}-\{j[(j-1+u)(1+c)+vc\\
    & +a+b+1-u-v]+q\}s_{j}+(j+1)(j+u)s_{j+1}=0,
  \end{split}
\end{align}
with the initial conditions
\begin{equation}
s_{0}=1,\quad s_{1}= \frac{q}{uc}, \quad \mbox{and}\quad  s_{j}=0, \quad \mbox{if}\quad  j<0.
\end{equation}
\noindent The solution Eq.(\ref{L16}) is valid only within a circle centered at the origin $\varphi=0$ whose radius is the distance from the origin to the nearest singularity $\varphi=1$ or $\varphi=c$. For $c>1$, the radius of convergence is 1[33]. From Eq.(\ref{L17}), we can say that Heun function remains the same with the exchange of the parameters $a$ and $b$.
\subsection{Degeneracy to the Hypergeometric Models}
It can be easily verified that the Heun equation Eq.(\ref{L10}) can be reduced to the Hypergeometric equation in several ways [33]. They are
\begin{subequations}\label{grp}
\begin{align}
c&= 1, \quad q= ab,\label{second}\\
w&= 0, \quad q=cab,\label{third}\\
c&= 0, \quad q=0.
\end{align}
\end{subequations}
Let us now consider the simplest case of $c=0, q=0$. Then for $a+b=0$ and $1/2 -v=-i\beta/2$,  Eq.(\ref{L10}) reduces to
standard form of the Gauss Hypergeometric equation
\begin{equation}\label{ghg}
\frac{d^{2}C_{a}}{d\varphi^{2}}+\left[\frac{r-(1+a+b)\varphi}{\varphi(1-\varphi)}\right]\frac{dC_{a}}{d\varphi}-
\frac{abC_{a}}{\varphi(1-\varphi)}=0.
\end{equation}
where $r=1/2-i\beta/2$. The general solution for Eq.(\ref{ghg}) is
\begin{equation}
\begin{split}
 C_{a}=\mbox{P}_{1}\varphi^{1-r}& \mbox{F}[b-r+1,a-r+1;2-r;\varphi] \\
& + \mbox{P}_{2}\mbox{F}[a,b;r;\varphi],
\end{split}
\end{equation}
where the constants, $\mbox{P}_{1} , \mbox{P}_{2}$ can be found using the initial conditions of the problem.  We write the hypergeometric series $\mbox{F}_{(2,1)}[a,b;c;\varphi]$ as F$[a,b;c;\varphi]$. The population left in the state $ \left |a \right \rangle $  is given as
\begin{equation}
C_{af}= \mbox{P}_{1}\mbox{F}[b-r+1,a-r+1;2-r;1]+ \mbox{P}_{2}\mbox{F}[a,b;r;1].
\end{equation}
Subsequently if $(a+b)=\lambda i\beta$ and $v-1/2-(a+b)=\mu i\beta$, we have the generalized Rosen-Zener Model as discussed by Bambini and Berman[21]. One can summarize the degeneracy of the Heun to Hypergeometric model as follows
\begin{subequations}\label{grp}
\begin{align}
\mbox{H}l\left[1,ab;a,b,u,v;\varphi\right]= \mbox{F}[a,b;u;\varphi], \label{second}\\
\mbox{H}l\left[c,cab;a,b,u,a+b-u+1;\varphi\right]=  \mbox{F}[a,b;u;\varphi], \label{third}\\
\mbox{H}l\left[0,0;a,b,u,v;\varphi\right]=  \mbox{F}[a,b;a+b-v+1;\varphi].
\end{align}
\end{subequations}
\begin{figure*}
\includegraphics[height=6cm,width=0.45\textwidth,angle=0]{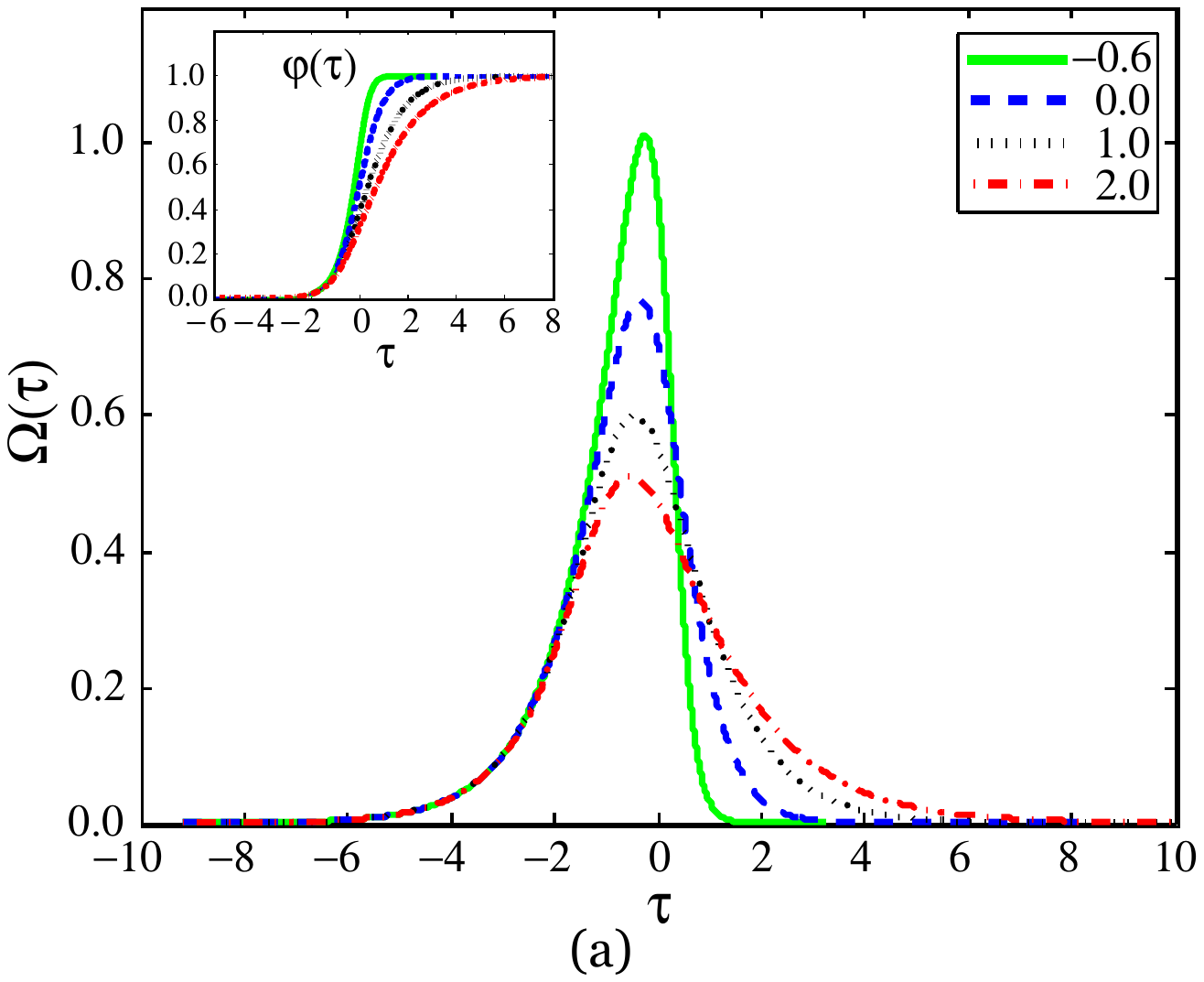}
\includegraphics[height=6cm,width=0.45\textwidth,angle=0]{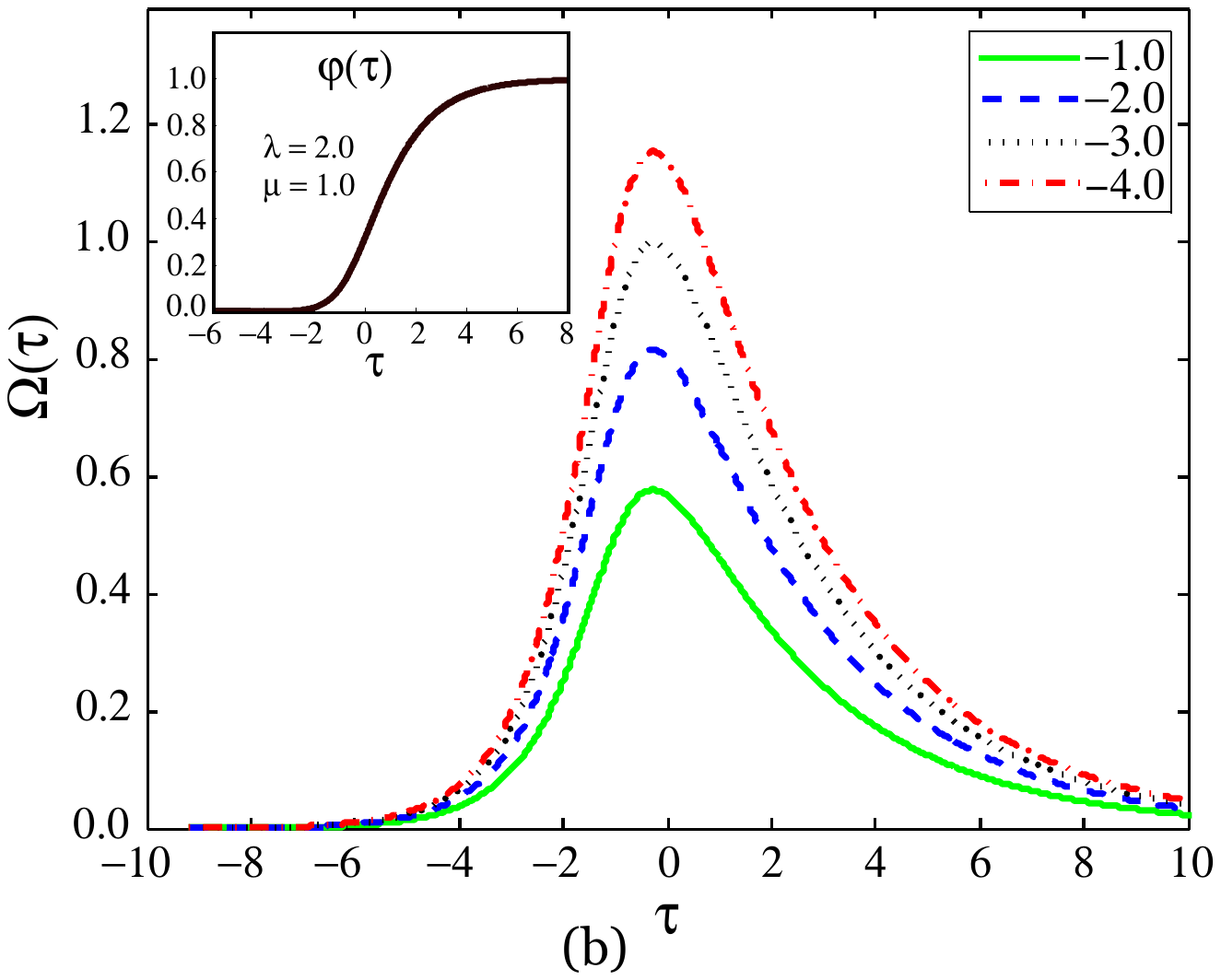}
  \caption{(Color online) Pulse shapes given by Eq(31).(a) Pulse shapes with varying $\lambda$ and $p=-q=1$. (b) Pulse shapes with varying $q$ and $\lambda=2, p=0.$}
\end{figure*}
\subsection{ Confluent Heun Equation}
 The Confluent Heun Equation is one of the four confluent forms of Heun's equation which is obtained by merging the singularity at $\varphi=c$ that at $\varphi=\infty$. Now we have a regular singularity at $\varphi=0,1$ and an irregular singularity at $\varphi=\infty$. In this paper we will consider the following non-symmetrical form of the Confluent Heun equation:
\begin{equation}\label{L19}
\frac{d^{2}C_{a}}{d\varphi^{2}}+\left(\frac{u}{\varphi}+\frac{v}{\varphi-1}\right)\frac{dC_{a}}{d\varphi}+
\frac{p\varphi+q}{\varphi(\varphi-1)}C_{a}=0.
\end{equation}
Similar to the Heun case, we have the same differential equation for $\dot{\varphi}$ i.e Eq(\ref{L13}). For the Confluent Heun Equation,  the possible values of the asymmetric parameters are
\begin{subequations}\label{grp}
\begin{align}
u&= \frac{1}{2}-\frac{i\beta\mu}{2},\quad v= \frac{i\beta(\lambda +\mu)}{2},\quad p=-q, \label{second}\\
u&= \frac{1}{2}-\frac{i\beta\mu}{2},\quad v= \frac{1}{2}+\frac{i\beta(\lambda +\mu)}{2},\quad p=0
\end{align}
\end{subequations}
The general solution of the Confluent Heun Equation Eq.(\ref{L19}) is given as
\begin{align}\label{L21}
  \begin{split}
    &C_{a}=\mbox{P}_{1}\mbox{H}l^{(c)}[0,u-1,v-1,p,q+(1-uv)/2,\varphi] \\
    &+ \mbox{P}_{2}\varphi^{1-u}\mbox{H}l^{(c)}[0,1-u,v-1,p,q+(1-uv)/2,\varphi],
  \end{split}
\end{align}
where $\mbox{P}_{1}$, $\mbox{P}_{2}$ can be found using the initial condition of the system. It is worth mentioning here that, the general solution to the Gauss Hypergeometric differential equation Eq.(\ref{ghg}) can be expressed
in terms of the Heun functions $\mbox{H}l^{(c)}$ as
\begin{align}
  \label{eq:nine}
  \begin{split}
    &C_{a}=\mbox{P}_{1}(\varphi -1)^{-a}\mbox{H}l^{(c)}[0,a-b,-1+r,0,((r-2a)b-r\\
    &+ra+1)/2,1/(1-\varphi)]  +\mbox{P}_{2}(\varphi -1)^{-b}\mbox{H}l^{(c)}[0,b-a,\\
    &-1+r,0,((r-2a)b-r+ra+1)/2,1/(1-\varphi)].
  \end{split}
\end{align}
The form of the pulse can be obtained by equating Eq.(\ref{L7}) and Eq.(\ref{L19}) which gives,
\begin{equation}\label{Pulse1}
\Omega(\varphi)=\frac{\left[4\varphi(\varphi-1)(p\varphi+q)\right]^{1/2}}{\mu+\lambda\varphi},
\end{equation}
where $\varphi(\tau)$ is given by Eq.(\ref{L14})[34].
The constraint of $\lambda$ and $\mu$ is also the same as for the Heun case discussed earlier. Fig.(3) shows the pulse shapes for which the two-level atom can be reduced to the Confluent Heun equation.
It also qualitatively shows the effect of the asymmetric parameters $p$ and $ q$ on the symmetry of the pulse shapes. $\lambda=0$ corresponds to the symmetric pulse.
\section{SOME EXAMPLES}
 In this section we will consider some specific examples of pulses corresponding to Heun and Confluent Heun equations. Interestingly we will also find a better approximation for a box pulse by introducing a parameter $\delta$ which takes care of non-analyticity of the pulse at the edges.
\subsection{$\Omega_{\delta}(t)=\Omega_{0} \mbox{sech}(\alpha t) / \sqrt{\delta-\mbox{tanh}(\alpha t)},\quad \delta > 1$}
\noindent For this pulse,
using the scaling parameters Eq.(\ref{L5}), Eq.(\ref{L4}) gives
\begin{align}\label{L22}
  \begin{split}
    \ddot{C}_{a}(\tau)-&\left[i\beta +\frac{1}{2}\left(\frac{1-2\delta \mbox{tanh}\tau + \mbox{tanh}^{2}\tau}{\delta - \mbox{tanh}\tau}\right)\right]\dot{C}_{a}(\tau) \\
    &+\frac{\gamma^{2}\mbox{sech}^{2}\tau }{\delta-\mbox{tanh}\tau}C_{a}(\tau)=0.
  \end{split}
\end{align}
\noindent Let us now define a new variable as
\begin{equation}\label{L23}
\varphi(\tau)=\frac{1+\mbox{tanh}\tau}{2}.
\end{equation}
In terms of the variable $\varphi$, Eq.(\ref{L22}) reduces to the Heun equation
\begin{equation}\label{L24}
C^{''}_{a}+\left[\frac{u}{\varphi}+\frac{v}{\varphi-1}+\frac{w}{\varphi-c}\right]C^{'}_{a}+\frac{ab\varphi-q}{\varphi (\varphi-1)(\varphi-c)}C_{a}=0,
\end{equation}
where,
\begin{subequations}\label{L25}
\begin{align}
u&= \frac{1}{2}-\frac{i\beta}{2},\quad v= \frac{1}{2}+\frac{i\beta}{2},\quad w= \frac{1}{2}, \label{second}\\
q&=-\frac{\gamma^{2}}{2},\quad a= 0, \quad b=\frac{1}{2}\quad c=\frac{\delta +1 }{2}.
\end{align}
\end{subequations}
From Eq.(\ref{L22}) we see as $\tau \rightarrow -\infty$, $\varphi \rightarrow 0$ and $\tau \rightarrow \infty$, $\varphi \rightarrow 1$. The initial conditions for our system are
\begin{equation}\label{L26}
C_{a}(\tau \rightarrow -\infty)=0, \quad |C_{b}(\tau \rightarrow -\infty)|=1.
\end{equation}
\noindent The complete solution to Eq.(\ref{L24}), satisfying the initial conditions Eq(\ref{L26}), is
\begin{align}
  \label{eq:two}
  \begin{split}
    &C_{a}(\varphi)=\frac{\gamma \sqrt{2}}{(i-b)\sqrt{c}}\varphi^{1-u}\mbox{H}l[c,q+(1-u)((c-1)v\\
    &+a+b-u+1); b-u+1,a-u+1,2-u,v,\varphi].
  \end{split}
\end{align}
where $a,b,c,q,u,v,w$ are given be Eq.(\ref{L25}).
\begin{figure}
  \includegraphics[height=4.5cm,width=4.2cm,angle=0]{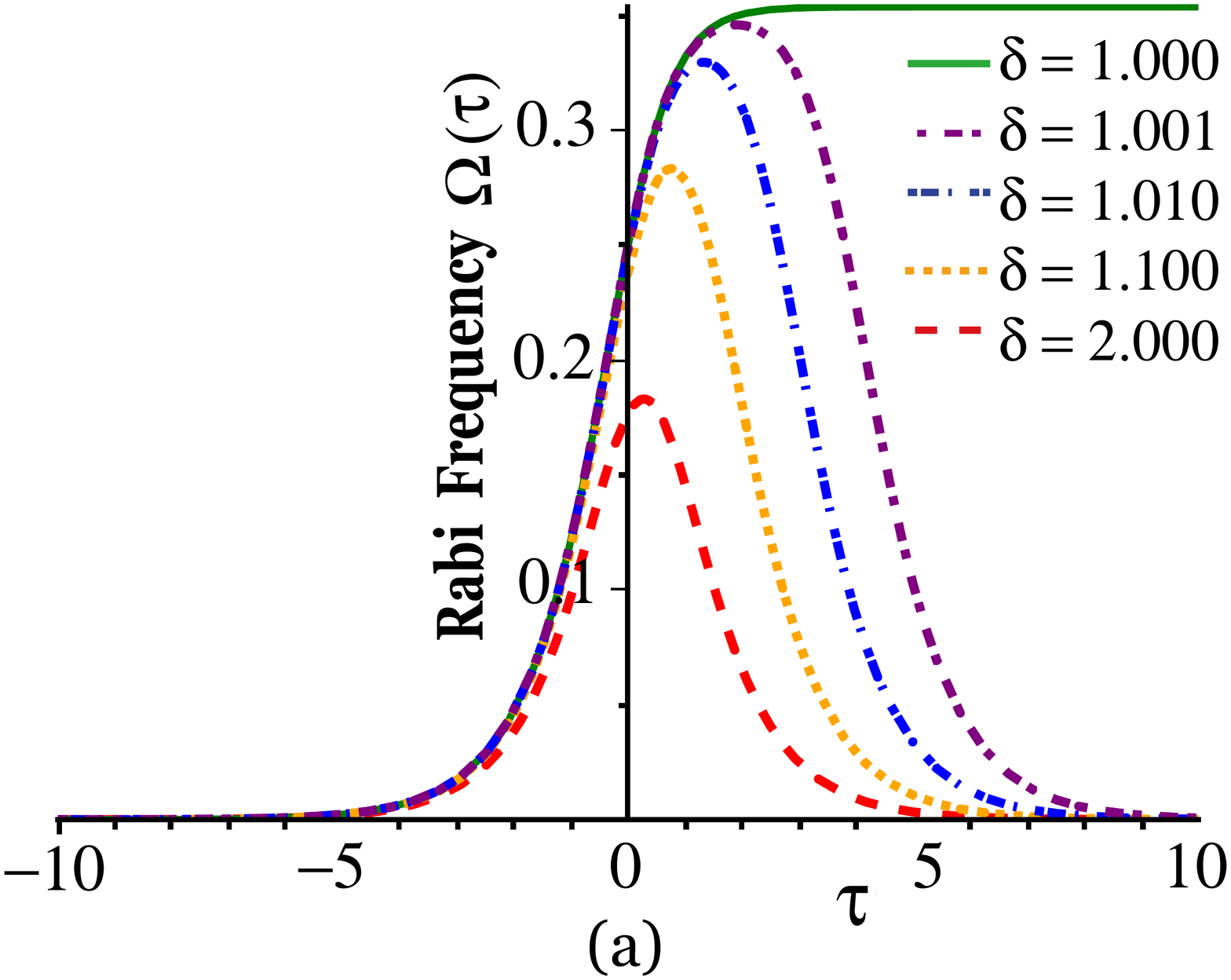}
  \includegraphics[height=4.5cm,width=4.3cm,angle=0]{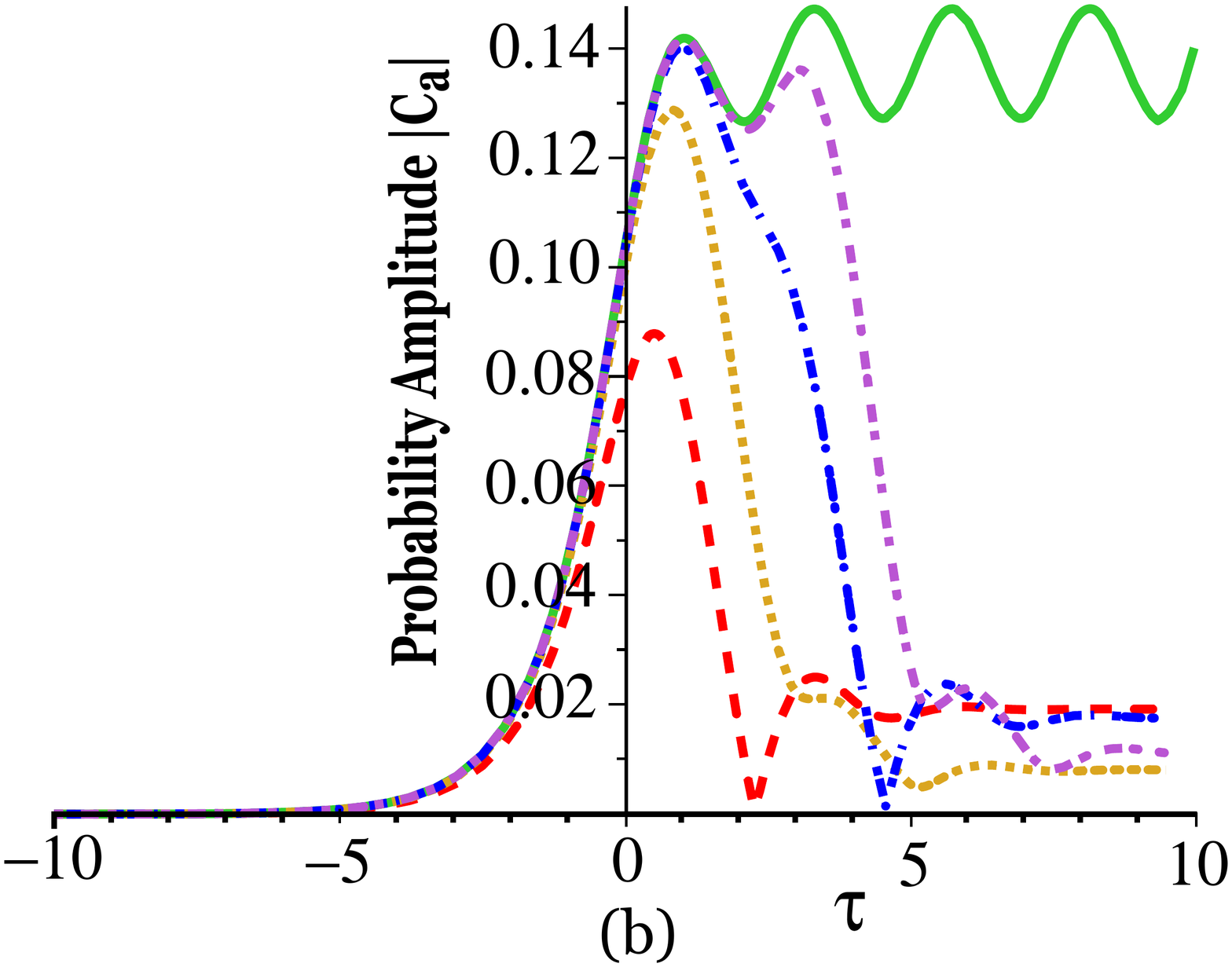}
 \caption{(Color online) (a) Pulse shapes for different value of $\delta$. (b) The time dependence of the population in the state $ \left |a \right \rangle $ for $\Omega_{\delta}(\tau)$ pulse for different values of $\delta>1$. For calculation we take $\alpha=0.08\omega_{c}$, varying $\delta$.}
\end{figure}
Let now consider a case in which $\delta=1$. So the pulse has the form
\begin{equation}
\Omega_{1}(t)=\Omega_{0}\sqrt{1+\mbox{tanh}\alpha t}.
\end{equation}
\noindent Now for this pulse,
using the scaling parameters Eq.(\ref{L5}), Eq.(\ref{L4}) gives
  \begin{equation}\label{L27}
    \ddot{C}_{a}(\tau)-\left[i\beta +\frac{1}{2}\left(1-\mbox{tanh}\tau\right)\right]\dot{C}_{a}(\tau)
    +\gamma^{2}(1+\mbox{tanh}\tau)C_{a}(\tau)=0.
\end{equation}
In terms of the variable $\varphi$, Eq.(\ref{L27}) reduces to
\begin{equation}\label{L28}
C^{''}_{a}+\left[\frac{u}{\varphi}+\frac{v}{\varphi-1}\right]C^{'}_{a}+\frac{q}{\varphi (\varphi-1)^{2}}C_{a}=0,
\end{equation}
where,
\begin{equation}\label{L29}
u= \frac{1}{2}-\frac{i\beta}{2},\quad v= 1+\frac{i\beta}{2},\quad  q=\frac{\gamma^{2}}{2}.
\end{equation}
\noindent The general solution to  Eq.(\ref{L28}) is
\begin{align}
  \begin{split}
    &C(\varphi)_{a}=\mbox{P}_{1}(\varphi-1)^{\xi} \mbox{F}[\xi,\xi-1+u+v;u;\varphi]       \\
    &\mbox{P}_{2}\varphi^{1-u}(\varphi-1)^{\xi} \mbox{F}[\xi+v,\xi+1-u;2-u;\varphi],
  \end{split}
\end{align}
where,
\begin{equation}
\xi=\frac{1-v}{2}+\sqrt{\left(\frac{1-v}{2}\right)^{2}-q},
\end{equation}
and $q,u,v$ are given be Eq.(\ref{L29}).
Using the initial conditions Eq.(\ref{L26}) we get $P_{1}=0$ and
\begin{equation}
\mbox{P}_{2}=\frac{\gamma}{\sqrt{2}(u-1)(-1)^{(\xi +1/2)}}
\end{equation}
Fig.(4) shows the plot of population in the state $ \left |a \right \rangle $ corresponding to the pulse $\Omega_{\delta}$ satisfying the initial condition.
\subsection{$\Omega_{+}(t)=\Omega_{0} \mbox{sech}\alpha t  (\sqrt{1+\mbox{tanh}\alpha t})$}
\noindent For this pulse, using the scaling parameters Eq.(\ref{L5}), Eq.(\ref{L4}) gives

\begin{align}\label{L30}
  \begin{split}
    &\ddot{C}_{a}(\tau)-\left[i\beta +\frac{1}{2}(1-3\mbox{tanh}\tau)\right]\dot{C}_{a}(\tau) \\
    & +\gamma^{2}\mbox{sech}^{2}\tau (1+\mbox{tanh}\tau) C_{a}(\tau)=0.
  \end{split}
\end{align}
In terms of the new variable $\varphi$, Eq.(\ref{L30}) reduces to the Confluent Heun equation.
\begin{equation}\label{L31}
C^{''}_{a}+\left[\frac{u}{\varphi}+\frac{v}{\varphi-1}\right]C^{'}_{a}+\frac{\sigma}{\varphi-1}C_{a}=0,
\end{equation}
where,
\begin{equation}
u= -\frac{i\beta}{2},\quad v= \frac{1}{2}+\frac{i\beta}{2}, \quad \sigma=-2\gamma^{2}.
\end{equation}
The complete solution to Eq.(\ref{L31}) satisfying the initial conditions Eq.(\ref{L26}) is
\begin{align}
  \begin{split}
    C_{a}(\varphi)=\left(\frac{2\sqrt{2}\gamma}{2i-\beta}\right)&\varphi^{1+\frac{i\beta}{2}}\mbox{H}l^{(c)}[0,1+i\beta/2,-1/2+i\beta/2, \\
    & -2\gamma^{2},1/2-\beta^{2}/8-i\beta/8,\varphi].
  \end{split}
\end{align}
\subsection{$\Omega_{-}(t)=\Omega_{0} \mbox{sech}\alpha t  (\sqrt{1-\mbox{tanh}\alpha t})$}
\noindent For this pulse, using the scaling transformation Eq.(\ref{L5}), Eq.(\ref{L4}) gives
\begin{align}\label{L32}
  \begin{split}
    &\ddot{C}_{a}(\tau)-\left[i\beta -\frac{1}{2}(1+3\mbox{tanh}\tau)\right]\dot{C}_{a}(\tau) \\
    & +\gamma^{2}\mbox{sech}^{2}\tau (1-\mbox{tanh}\tau) C_{a}(\tau)=0.
  \end{split}
\end{align}
In terms of the new variable $\varphi$, Eq.(\ref{L32}) reduces to the Confluent Heun equation.
\begin{equation}\label{L33}
C^{''}_{a}+\left[\frac{u}{\varphi}+\frac{v}{\varphi-1}\right]C^{'}_{a}+\frac{\eta}{\varphi}C_{a}=0,
\end{equation}
where,
\begin{equation}
u= \frac{1}{2}-\frac{i\beta}{2},\quad v= \frac{i\beta}{2}, \quad \eta=2\gamma^{2}.
\end{equation}
The complete solution to Eq.(\ref{L33}), satisfying the initial conditions Eq.(\ref{L26}), is
\begin{align}
  \begin{split}
C_{a}(\varphi)=\left(\frac{2\sqrt{2}\gamma}{\beta-i}\right)&\varphi^{\frac{1}{2}+\frac{i\beta}{2}}\mbox{H}l^{(c)}[0,1/2+i\beta/2,-1+i\beta/2,2\gamma^{2}, \\
    & 1/2-2\gamma^{2}-\beta^{2}/8-i\beta/8,\varphi].
  \end{split}
\end{align}
\begin{figure}[htb]
  \includegraphics[height=4.5cm,width=4cm,angle=0]{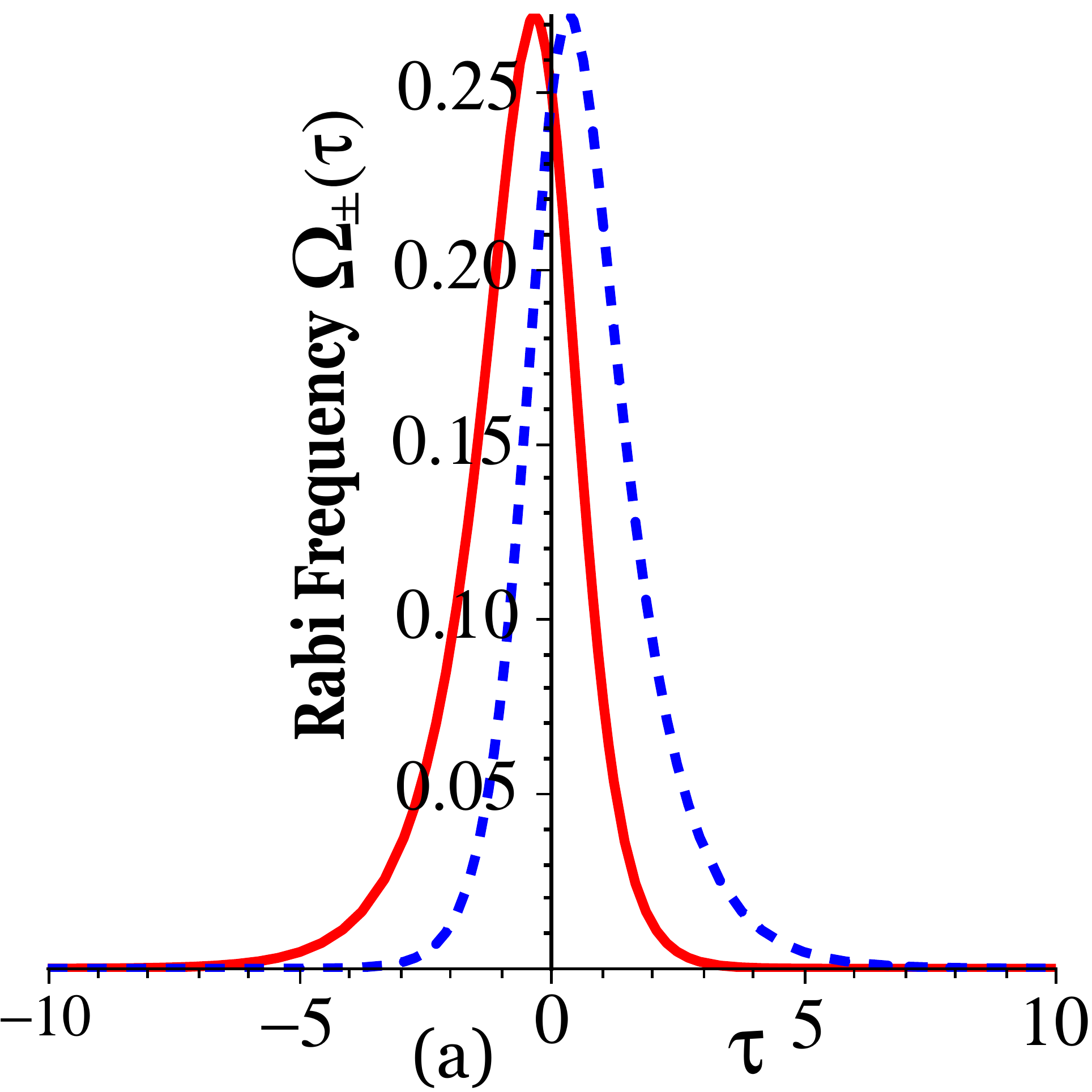}
  \includegraphics[height=4.5cm,width=4cm,angle=0]{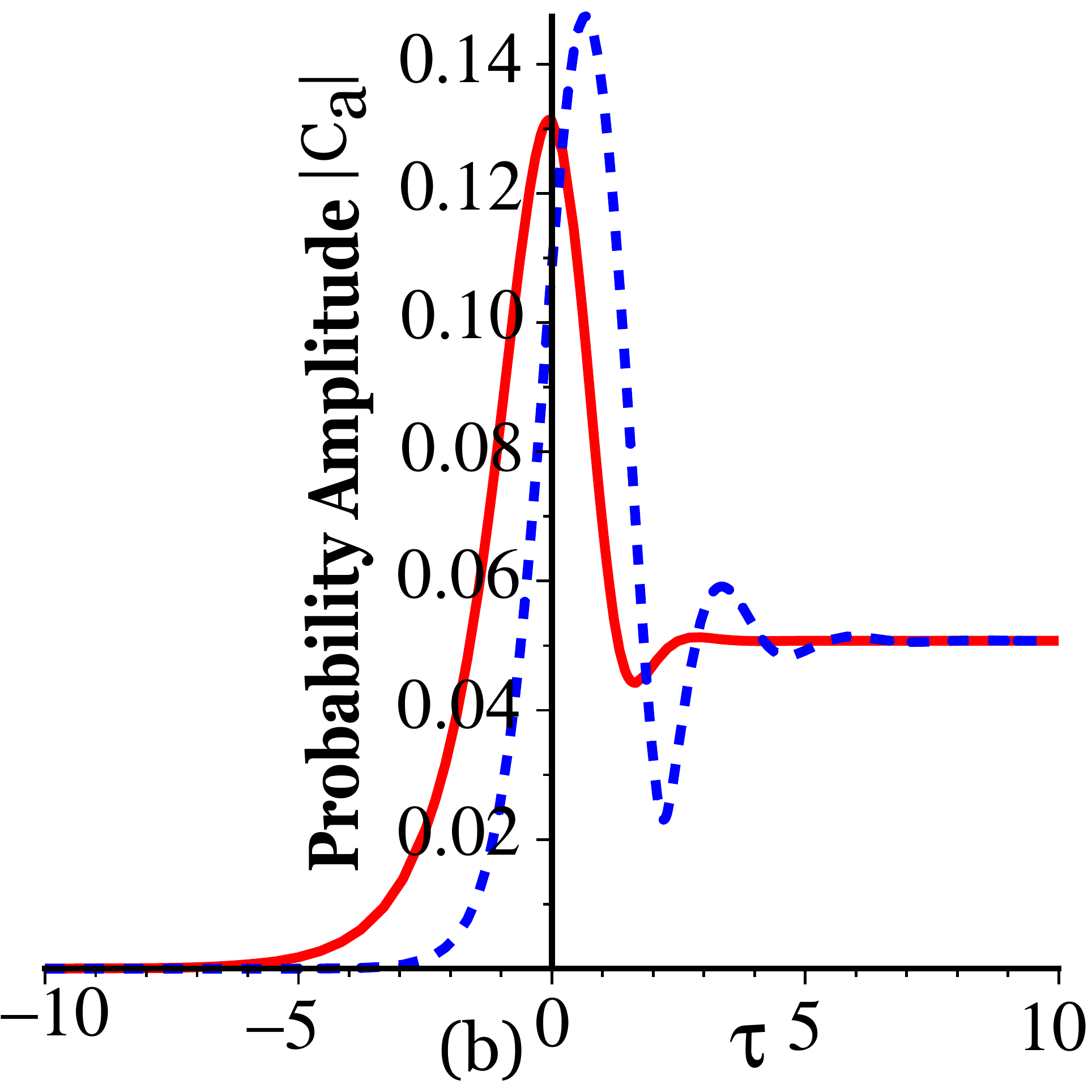}
 \caption{(Color online) (a) Pulse shapes for $\Omega_{\pm}(t)=\Omega_{0} \mbox{sech}\alpha t  (\sqrt{1\pm \mbox{tanh}\alpha t})$. (b) Time dependence of population in the state $ \left |a \right \rangle $ for the Pulse shapes in(a). In calculation we take $ \Omega_{0}=0.02\omega_{c} , \alpha=0.08\omega_{c}, \Delta=0.2\omega_{c}$ }
\end{figure}
In Fig(5) we have plotted the pulse shapes $\Omega_{\pm}(\tau)$ and the corresponding time evolution of the probability amplitude for state $|a\rangle$.
\subsection{Smooth Box Pulse}
One of the simplest and exactly solvable pulse shapes is a Box Pulse. Indeed it is a non-analytical pulse but it gives information about the basic oscillatory nature of solution (probability amplitude). Let us define our pulse as
\begin{equation}
\Omega(t)=\Omega_{0}\Theta(t)\Theta(t_{0}-t), \quad t_{0}>0
\end{equation}
where, $\Theta(t)$ is a unit step function. The solution for Eq.(\ref{L4}) corresponding to the box pulse is
\begin{equation}
C_{a}(t)=\frac{i\Omega_{0}}{\sqrt{\Delta^{2}/4 +\Omega_{0}^{2}}}e^{i(\Delta/2)t} \mbox{sin}(\sqrt{\Delta^{2}/4 +\Omega_{0}^{2}})t, \quad t<t_{0}
\end{equation}
The oscillatory nature of the solution $| C(t) |$ is evident from the sine function.
Let us consider the pulse shape of the form
\begin{equation}\label{L34}
\Omega_{\delta}(t)=\frac{\Omega_{0} \mbox{sech}\alpha t }{ \sqrt{\delta-\mbox{tanh}\alpha t}}, \quad \delta=2c-1
\end{equation}
\begin{figure}[b]
 \includegraphics[height=3.9cm,width=8.6cm,angle=0]{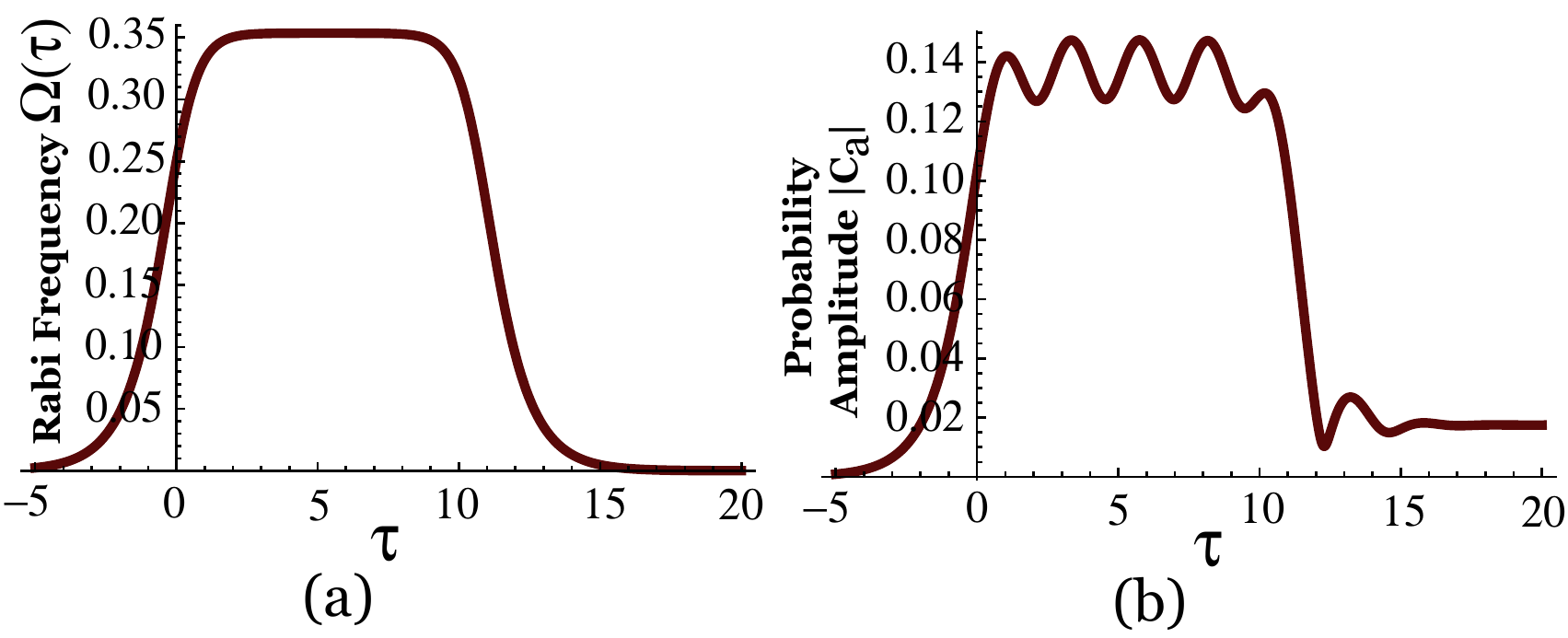}
 \caption{(a) Box Pulse for $\delta-1=10^{-9}$.  (b) Time dependence of population in the state $ \left |a \right \rangle $ for the Box Pulse $\Omega_{\delta}(\tau)$. In calculation we take $ \Omega_{0}=0.02\omega_{c} , \alpha=0.08\omega_{c}, \Delta=0.2\omega_{c}$ }
\end{figure}
where c is one of the singularities of the Heun Equation. Assuming $c > 1$ gives $\delta > 1$. A pulse shape of the form Eq.(\ref{L34}) is positive definite and it vanishes at $\tau=\pm\infty$. Let us see what happens when $\delta$ approaches but never reaches to 1. We see from Fig(4a),  that as $\delta$ approaches to 1, the pulse become more and more broad there by making it a better approximation for a box pulse (taking care of non-analyticity at the edges). The general solution for the pulse of the form Eq.(\ref{L34}), is given by Eq.(\ref{L15}) where the asymmetric parameters are given by Eq.(\ref{L25}).
\section{DISCUSSION}
\begin{figure}[b]
  \includegraphics[height=6cm,width=6cm,angle=0]{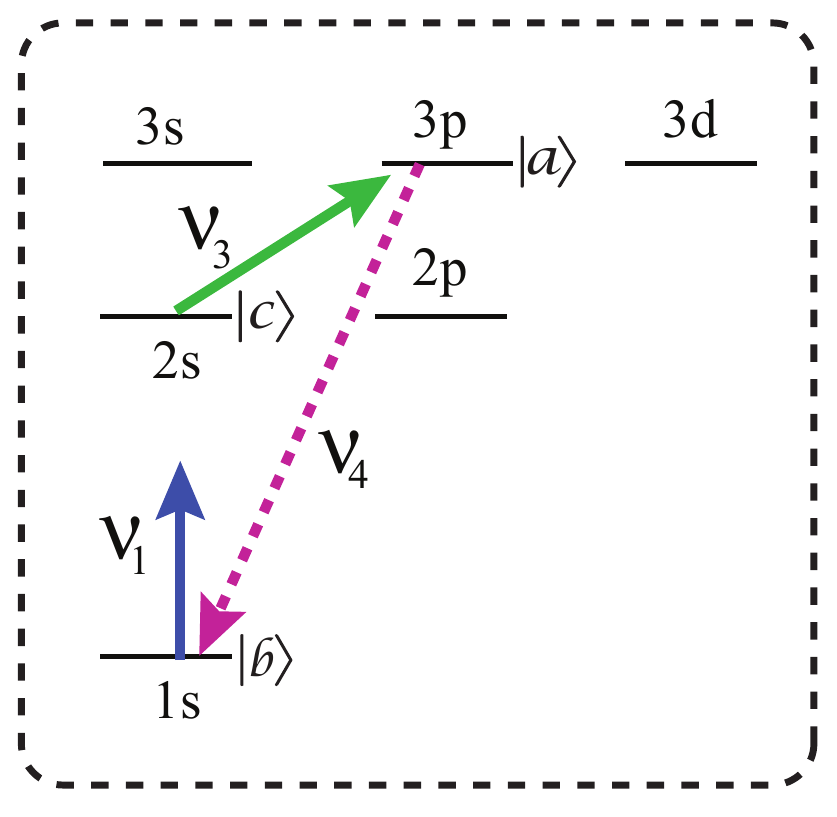}
  \caption{ (Color online) Field configuration and level structure of H or He$^+$.
All population is initially in the ground state $\left |b \right \rangle$.
First, the strong short far-off resonant pulse with frequency $\nu_1$ is applied to the system to
excite coherence between levels $\left |b \right \rangle=\left |1s \right \rangle$ and $\left |c \right \rangle=\left |2s \right \rangle$,
and then the second pulse with the frequency $\nu_3$, which is close to
the transition between levels $2s$ and $2p$, is applied to generate XUV pulse with higher frequency $\nu_4$}
\end{figure}
The obtained results can be applied to the generation of X-ray and UV (XUV) radiation, which is one of the main topics in modern optoelectronics and photonics~\cite{xuv}. Recent
progress in ultrashort, e.g. attosecond, laser technology allows searchers to obtain ultra-strong fields~\cite{strong-lasers}. Interaction of such strong and broadband fields with a two-level
atomic system, even under the action of a far-off resonance laser radiation is of current  interest~\cite{rost07jmo,mos08jmo,rost08jmo,rost09jmo,rost09pra}.  Strong short laser pulses can excite remarkable coherence on high frequency transitions; and this coherence can be used for surprisingly efficient generation of XUV radiation~\cite{mos08jmo,rost08jmo,rost09jmo,rost09pra}. In the first step we excite the atoms (e.g., from the 1s to 2s
states of or He$^+$, etc.) via a short pulse of femto- or attosecond radiation e.g., from a conventional Ti-sapphire laser system). The excitation occurs due to the coherent coupling between 1s and 2p and then 2p and 2s. In
the second step, we apply another pulse which scatters off the Raman coherence (prepared in the first step), generating short wavelength anti-Stoke radiation as depicted in Fig(7). The generation of radiation is a coherent process that (contrary to conventional superfluorescence) does not require population inversion (see Appendix). The higher efficiency of coherent process  has been demonstrated in various spectral regions[43-50].

We have analytically calculated above that the level of excited coherence when a two level atom is driven by a ultra-short intense pulse. The coherence is sufficiently large that this can be used for nonlinear generation of XUV radiation, i.e, see Figs(4b, 5b, 6b), coherence can be of the order of $0.1$.  It is instructive to estimate the level of XUV field that can be generated by using this coherence. After an ultra-strong and short pulse, we apply a strong resonant and relatively long pulse. The applied probe pulse $\W_3$ and  generated signal $\W_4$ are coupled to each other via
coherence excited in the medium (Rabi frequencies are defined as
$\W_{3,4} = \wp_{3,4} E_{3,4}/\hbar$).
Hence, the propagation equation for $\W_4$ is
given by
\begin{figure}[t]
 \includegraphics[height=5cm,width=0.5\textwidth,angle=0]{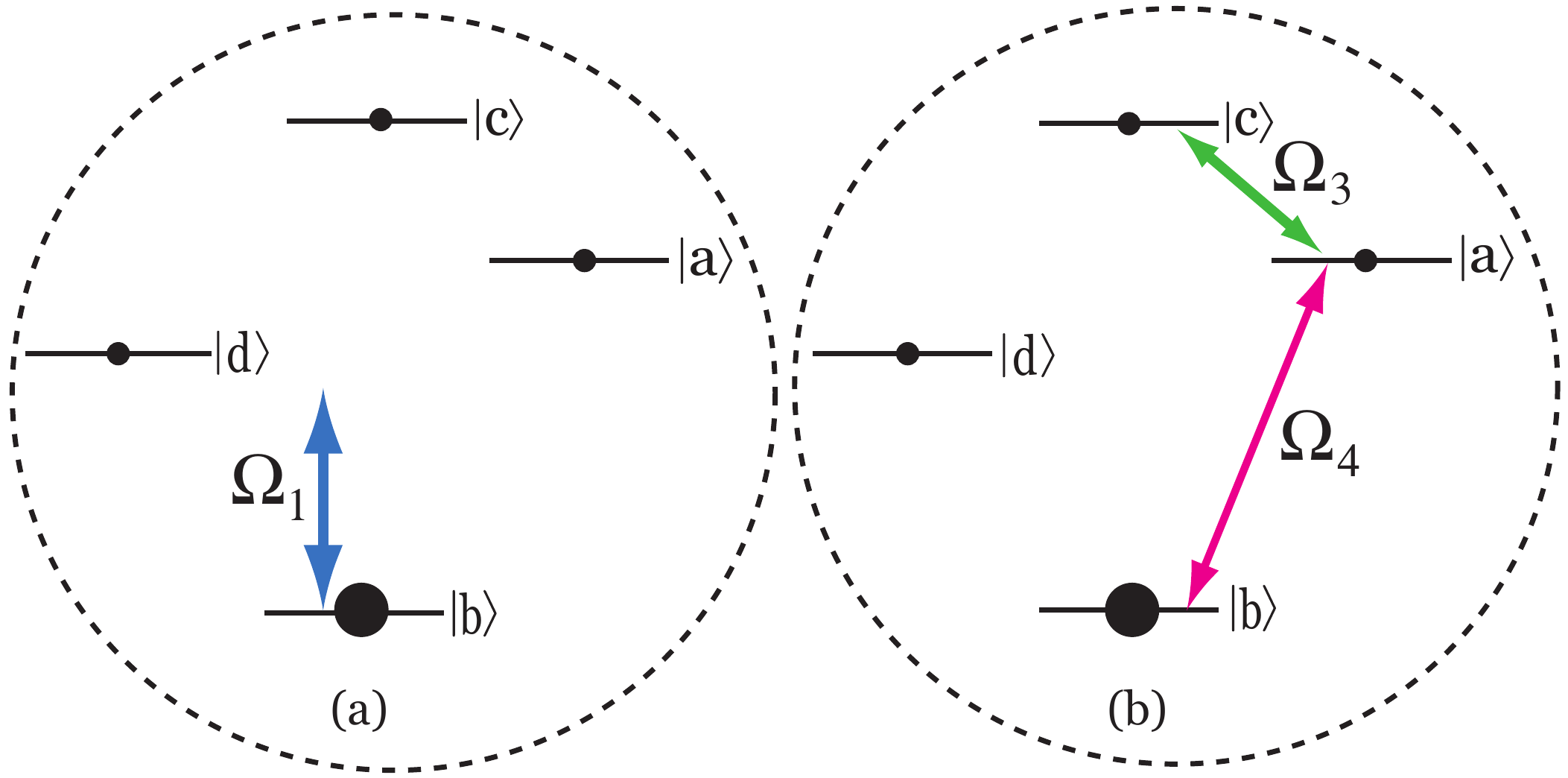}
\caption{(Color online) Two-stage generation scheme for X-ray generation. (a) Applying a strong pulse allows one to excite an atomic system by transferring population to electronic excited states. (b) Coherence is then induced by applying a resonant field. }
\end{figure}
\be
{\partial\W_4\over\partial z} = -i\eta_4\rho_{ab},
\ee
where $\rho_{ab}$ is the appropriate atomic coherence (see Fig.(8)),
and $\eta_{4} = k_{4} \wp^2_{4} N/2\hbar$, where $\wp_4$ is the dipole
moment at the transition between .
The corresponding equation for the density matrix coherence $\rho_{ab}$ is
\be
\dot\rho_{ab} = -\G_{ab}\rho_{ab} + i\W_4(\rho_{aa} - \rho_{bb}) -i\W_3\rho_{cb},
\ee
and, for short pulses, $\rho_{ab} \simeq -i\W_3\tau\rho_{cb}$.
Then, we can estimate the intensity of the signal field, by
\be
\W_4 = {k_4 L \wp_{ab}^2N\over2\hbar} \rho_{cb} \W_3\tau,
\ee
where $k_4$ is the wavenumber for signal radiation, $L$ is the length of the
active medium, $\wp_{ab}$ is the dipole moment at the transition between $a$
and $b$ levels, $\tau$ is the time duration of the pump laser pulse.
Using the parameters $N\simeq 10^{16-19}$ cm$^{-3}$, $\wp_{ab}\simeq 1D$,
$L=100$ $\mu$m, $\rho_{cb} = 10^{-1}$, $\W_3\tau = 1-10^3$,
$\tau = 1$ ps, $\lambda = 10$ nm, we obtain
$\mbox{energy} \simeq 10\; nJ - 1\; \mu J$,
This estimate shows the promise of the approach. This estimate is valid
on the time scale when the collisions in the plasma destroy the coherence.
It occurs at the times of order
$\delta t = 1/\sigma c N \simeq 1~\mathrm{ps}$, where
$\s$ is the atomic cross-section for atomic collisions that destroy the excited
coherence.
\section{Conclusion}
In this paper we have found several analytical solutions for a two-level atomic system under the action of a far-off resonance strong pulse of laser radiation.  The solutions are given in terms of  Heun function which is a generalization of the Hypergeometric function. The Rosen-Zener and Bambini-Berman Model belongs to this class of pulses as  special cases. A better approximation for box pulse is also obtained here which take care of non-analyticity at the edges by introducing a parameter $\delta$. The results obtained here have applications to the generation of XUV radiation and the estimate shown above shows a good potential for a source of coherent radiation. The technique used here to get the exactly solvable pulse shapes can be generalized to appropriate time dependent detuning $\Delta=\Delta(t)$ cases and produce more exactly solvable models (to be reported elsewhere)[51]
\section{Acknowledgements}
We thank M.O.Scully, L. Keldysh, M.S.Zubairy, V. A. Sautenkov, H.Eleuch, A. Svidzinsky, H. Li and E. Sete  for useful discussions and gratefully acknowledge the support from the NSF Grant EEC-0540832 (MIRTHE ERC),the Defense Advanced Research Projects, Office of Naval Research (N00014-07-1-1084 and N0001408-1-0948), Robert A. Welch Foundation (Award A-1261)) and the partial support from the CRDF. P.K.Jha would also like to acknowledge the Robert A. Welch Foundation Graduate Fellowship.

\appendix
\section{Generation of radiation by a two-level atomic medium with excited coherence}
Let us assume that a two-level atom has some small initial coherence
$\rho_{ab}^0 = \sqrt{\rho_{aa}^0\rho_{bb}^0}$. Note that in this paper,
we consider the case when there is no
population inversion, $\rho_{aa}^0 < \rho_{bb}^0$.
The density matrix equations for atomic coherence are
\be
{\partial\rho_{ab}
\over\partial t} = i\W(\rho_{aa}-\rho_{bb}),\mbox{ and }
\ee\be
{\partial
\over\partial t}(\rho_{aa} - \rho_{bb}) = -2i\W\rho_{ab}.
\ee
the solution (by neglecting relaxation processes) is
\be
\rho_{ab} = i\rho_{ab}^0\sin\theta.
\ee
Then, for the retarded frame
\be
\tau = t - {z\over c},
\ee
the propagation equation for a resonant field is given by
\be
{\partial\W
\over\partial z} = -i\eta\rho_{ab},
\label{W-theta}
\ee
where $\eta=3\lambda^2 N\g/(8\pi)$ is the coupling constant.
Introducing
\be
\theta = 2\int^t \W\; dt,
\ee
Eq.(\ref{W-theta}) can be rewritten as
\be
{\partial^2\theta
\over\partial z\partial\tau} = -\eta\sin(\theta - \phi),
\label{W-theta2}
\ee
where $\phi$ can be determined from initial condition as
\be
\phi \simeq 2\sqrt{\rho_{aa}^0}.
\ee
Solution of Eq.(\ref{W-theta2}) is given by
\be
\theta = \phi[1 - J_0(2\sqrt{\eta z\tau})],
\ee
and the Rabi frequency is
\be
\W = \phi J_1(2\sqrt{\eta z\tau})\sqrt{\eta z\over \tau}.
\ee
The energy of the generated short wavelength pulse can be calculated
as
\beq
{c\over4\pi}A \int^{\infty}_{-\infty} |E|^2 dt =
Az \;N\;\rho_{aa}\;\;\hbar\w_{ab},
\eeq
and it is equal to the energy stored in the medium after excitation.
Also it is important to note that the absence of population inversion
does not influence much of pulse energy because of coherent interaction
of the radiation field with the atomic medium.

The time duration of the generated pulse is of the order of
\be
\tau_{pulse} = {4\pi\over 3N\lambda^2 z \gamma_r},
\ee
and it gives the power of the pulse be
\be
P_{pulse} = {\lambda^2 z N\over 4\pi}Az N \gamma_r \rho_{aa} \hbar\omega_{ab},
\ee
where the factor $\ds{\lambda^2 z N\over 4\pi}$ shows
the brightness of the source in comparison with spontaneous emission of incoherent source.


\begin{thebibliography}{99}
\bibitem{PC1} In this paper we have defined the Rabi frequency $\Omega(t)=\wp{\cal E}(t)/2\hbar$ rather than the usual definition $\Omega(t)=\wp{\cal E}(t)/\hbar$.
\bibitem{PC2}Here we consider a two-level atom with stable levels (or neglect any kinds of decay due to spontaneous emission, collision etc on the time scale of the pulse) interacting with a classical external electromagnetic field.
\bibitem{PC3}Heun Equation: For real $\Omega(\tau)$, we get an additional constraint  for our asymmetric parameters
$q<0,  ab/q < 1,\text{if}\hspace{1mm} q\ne 0, \text{or}\hspace{1mm} ab > 0 , \text{if}\hspace{1mm} q=0$.
Confluent Heun Equation:
$q<0,  p/q >-1, \text{if }\hspace{1mm} q\ne 0, \text{or}\hspace{1mm} p< 0 ,  \text{if }\hspace{1mm} q=0.$
\end{thebibliography}
\end{document}